\documentclass[letter,twocolumn]{jpsj3}
\usepackage{txfonts}

\title{
Network-Growth Rule Dependence of Fractal Dimension of Percolation Cluster on Square Lattice
}

\author{
\name{Shu \surname{Tanaka}}
and
\name{Ryo \surname{Tamura}}$^{1,2}$
}
\inst{
Department of Chemistry, University of Tokyo,
\address{
7-3-1, Hongo, Bunkyo-ku, Tokyo, 113-0033, Japan
}\\
$^1$ Institute for Solid State Physics, University of Tokyo,
\address{
5-1-5, Kashiwanoha, Kashiwa-shi, Chiba, 277-8581, Japan
}\\
$^2$ International Center for Young Scientists, National Institute for Materials Science,
\address{
1-2-1, Sengen, Tsukuba-shi, Ibaraki, 305-0047, Japan
}
}
\abst{
To investigate the network-growth rule dependence of certain geometric aspects of percolation clusters, we propose a generalized network-growth rule introducing a generalized parameter $q$ and we study the time evolution of the network.
The rule we propose includes a rule in which elements are randomly connected step by step and the rule recently proposed by Achlioptas {\it et al.} [Science {\bf 323} (2009) 1453].
We consider the $q$-dependence of the dynamics of the number of elements in the largest cluster.
As $q$ increases, the percolation step is delayed.
Moreover, we also study the $q$-dependence of the roughness and the fractal dimension of the percolation cluster.
}

\kword{percolation, roughness, fractal dimension, network-growth model, Achlioptas rule}

\begin{document}
\maketitle

The study on percolation transitions has been an active topic of research not only in statistical physics but also in several areas of science such as materials science and information science\cite{Kirkpatrick-1973,Stauffer-1979,Stauffer-book}.
In materials science, percolation theory has been applied to investigate the relation between the connectivity of atoms and the properties of materials {\it e.g.} electric conductivity in alloys.
Dynamical phenomena such as spreading epidemics can be considered from the viewpoint of percolation transitions.
In information science, percolation theory can be also used to study the dynamic nature of evolving network systems such as the world wide web\cite{Albert-2002,Dorogovtsev-2008}.
It is believed that usual percolation transitions in static networks are continuous.
In percolation theory, the relation between the spatial dimensions and the critical exponents of percolation transitions has been established for static networks.
In dynamically evolving networks, however, to investigate the network-growth rule dependence of percolation transitions has been an important issue.

%A curious example is to investigate relationship between percolation phenomena and PageRank which is an algorithm for deciding importance of website and is used in search engine\cite{Frahm-2011}.
%Applicable scope of percolation theory has been extended day by day.

To investigate dynamically evolving networks, the simplest rule is as follows.
First, we randomly select two elements which are not connected.
Next, we connect the selected two elements deterministically.
After that, we repeat the above procedure.
Hereafter we refer to this rule as the random rule.
Under this rule, a continuous percolation transition occurs, which was shown by Erd\"os and R\'enyi\cite{Erdos-1960}.
However Achlioptas {\it et al.} proposed a new network-growth rule and studied percolation transitions under it\cite{Achlioptas-2009}.
They concluded that the percolation transition on a random graph resulting from their rule is discontinuous.
The rule proposed by Achlioptas {\it et al.} is as follows.
We first randomly select two pairs of elements which are not connected.
Next we compare the product of the number of elements in the cluster they belong to.
Then, we connect the elements with the smallest product.
If both products are equal, one of the two pairs is connected with equal probability.
In this paper, we call this the Achlioptas rule.
They studied the time evolution of the number of elements in the largest cluster, $n_{\rm max}$.
They calculated steps when $n_{\rm max}$ is equal to $\sqrt{N}$ and $N/2$, where $N$ is the number of elements in the network.
They numerically showed that steps in which these thresholds are crossed in the limit of $N \to \infty$ and concluded that the percolation transition is discontinuous.
They also mentioned that if we compare the sum of the number of elements instead of the product of the number of elements, the situation is qualitatively the same.
Since $n_{\rm max}$ explosively increases at the percolation point resulting from the Achlioptas rule, this percolation phenomenon is called ``explosive percolation''.
This study has attracted attention since discontinuous percolation transitions appear rarely.

After the study by Achlioptas {\it et al.}, a great amount of research has been devoted to the investigation of various aspects of explosive percolation.
Some of it is focused on network growth using the Achlioptas rule.
Ziff considered network growth on a square lattice under the Achlioptas rule and found that the percolation transition is discontinuous\cite{Ziff-2009,Ziff-2010}.
Cho {\it et al.} studied percolation transitions in scale-free networks under the Achlioptas rule\cite{Cho-2009}.
They found that explosive percolation is not always discontinuous and the continuity of the explosive percolation transition depends on the degree distribution of the scale-free network.
However, in some studies, the authors concluded that the explosive percolation is actually continuous\cite{Riordan-2011,Costa-2010,Grassberger-2011,Chae-2012}.
da Costa {\it et al.} proposed a representative model in which an explosive percolation occurs and insisted that the explosive percolation is continuous\cite{Costa-2010}.
Grassberger {\it et al.} obtained the probability distribution of $n_{\rm max}$ at the percolation point for a square lattice.
The distribution function is a bimodal distribution for finite size systems.
However, the distance between the two peaks decays with a power-law against the system size.
Then, the authors concluded that the explosive percolation is continuous\cite{Grassberger-2011}.
Moreover, several authors proposed generalized Achlioptas rules and considered the corresponding percolation transitions\cite{Araujo-2010,Cho-2011,Chen-2011,Araujo-2011,Fan-2012,Liu-2012}.
For example, Fan {\it et al.} introduced a probability $p$ which is related to the network-growth speed, and considered the network growth on a random graph\cite{Fan-2012}.
In their rule, two pairs of elements are randomly selected and the product of the number of elements in each cluster is compared as in the Achlioptas rule.
At each step the pair in which the product is the smallest is connected with probability $p$.
They analyzed the ratio of the number of elements in the second largest cluster divided by that in the largest cluster.
Then, they found that the percolation transition for $1/2 \le p \le 1$ is continuous.
As described above, it is still an open problem whether explosive percolation transitions are discontinuous but many interesting aspects of explosive percolation have been found\cite{Friedman-2009,Radicchi-2009,Cho-2010,Radicchi-2010,Ziff-2010,Lee-2011,Andrade-2011,Bashan-2011,Nagler-2011,Choi-2012,Nagler-2012}.

In this study, to explore the microscopic mechanism of such phenomena, we focus on the geometric aspects of percolation clusters.
To consider the dependence on network growth rules in a unified way, we propose a new generalized network-growth rule which includes both the random rule and the Achlioptas rule by introducing a generalized parameter $q$.
In addition, by adjusting $q$, we can construct a rule where the percolation step becomes fast in comparison to the random rule.
In our rule, we select two pairs of clusters and compare the sums of the number of elements, and one of them is connected probabilistically depending on the generalized parameter $q$.
Under the Achlioptas rule, it is a common occurrence that a cluster having a small number of elements is formed.
Furthermore, if we compare sums of numbers of elements, the number of elements obeys additivity both before and after the elements are connected by definition.
Then, we can clearly figure out what happens when we change the way in which we make network evolve.
Here, in order to graphically understand the geometric aspects of percolation clusters,
we consider network growth on a square lattice as the simplest case of regular graphs.

Before we explain our rule, we define the network-growth model and its notations.
Let $V$ and $E$ be a set of elements and edges, respectively.
The $i$-th ($1 \le i \le N$) element is denoted by $v_i \in V$, and $e_{ij} \in E$ represents the edge between $v_i$ and $v_j$ ($i \neq j$).
The state of $v_i$ is represented by $\sigma_i$ ($\sigma_i = \{1,\cdots, N\}$).
The state of edge $e_{ij}$ is expressed by $\tau_{ij}$ ($\tau_{ij} = \{0,1\}$).
%Edge $e_{ij}$ is said to be connected if $\sigma_i = \sigma_j$.
When the edge $e_{ij}$ is not connected, $\tau_{ij}=0$, whereas when the edge $e_{ij}$ is connected, $\tau_{ij}=1$.
A cluster is defined as a set of elements which are in the same state.
The number of elements in the cluster where $v_i$ belongs to is expressed by $n(\sigma_i)$.

\begin{figure}[t]
 \begin{center}
  \includegraphics[scale=0.75]{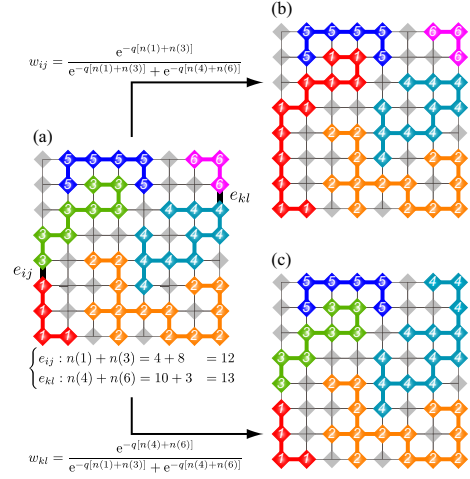}
  \caption{
  (Color online) Graphical representation of our proposed rule on a square lattice.
  The numbers in diamonds represent the state of the elements.
  The transparent gray diamonds denote the isolated elements.
  (a) The bold lines denote the randomly chosen edges $e_{ij}$ and $e_{kl}$.
  In this case, the sums of the clusters are $n(1) + n(3)=12$ and $n(4) + n(6) = 13$.
  We select $e_{ij}$(resp. $e_{kl}$) as the connecting edge with probability $w_{ij}$(resp. $w_{kl}$).
  In the random rule ($q=0$), we select $e_{ij}$ or $e_{kl}$ with the same probability $1/2$.
  In the Achlioptas rule ($q=+\infty$), we deterministically select $e_{ij}$ whereas we select $e_{kl}$ in the inverse Achlioptas rule ($q=-\infty$).
  (b) Configuration after the edge $e_{ij}$ is connected.
  (c) Configuration after the edge $e_{kl}$ is connected.
  }
  \label{fig:graphicalrep}
 \end{center}
\end{figure}

The procedure of our proposed network-growth rule is as follows:
\begin{description}
 \item[Step 1] The initial state is set to be $\sigma_i = i$ ($i=1,\cdots, N$) $\forall i$, in other words, all elements belong to different clusters, {\it i.e.} $n(\sigma_i)=1$ $\forall i$ and $\tau_{ij}=0$ $\forall i,j$.
 \item[Step 2] We randomly choose two different edges $e_{ij}$ and $e_{kl}$ such that $\tau_{ij} = \tau_{kl} = 0$, $\sigma_i \neq \sigma_j$, and $\sigma_k \neq \sigma_l$.
 \item[Step 3] Using a real number $q$, we connect $e_{ij}$ with the probability $w_{ij}$ defined by
\begin{eqnarray}
 \label{eq:weight}
 w_{ij} := \frac{
  {\rm e}^{-q\left[ n(\sigma_i) + n(\sigma_j) \right]}
}{
{\rm e}^{-q\left[ n(\sigma_i) + n(\sigma_j) \right]} +
{\rm e}^{-q\left[ n(\sigma_k) + n(\sigma_l) \right]}
},
\end{eqnarray}
whereas we connect $e_{kl}$ with the probability $w_{kl}:=1-w_{ij}$.
It should be noted that the treatment is similar to but differs from the rule proposed by Fan {\it et al.}\cite{Fan-2012}, since the probability $w_{ij}$ changes in each step depending on the shapes of the clusters.
After we connect $e_{ij}$(resp. $e_{kl}$) {\it i.e.} $\tau_{ij}=1$(resp. $\tau_{kl}=1$), the states of one of the clusters where $v_j$(resp. $v_l$) belongs to are changed to $\sigma_i$(resp. $\sigma_k$).
When we connect an edge, time advances from $T$ to $T+1$.
The introduced parameter $q$ in Eq.~(\ref{eq:weight}) is a generalized parameter and it characterizes the network-growth rule.
 \item[Step 4] We repeat steps 2 and 3 until all of the elements belong to the same cluster.
\end{description}
%Here we assume that all clusters are never separated.
In this network-growth rule, the number of clusters decreases one at a time.
In fact, the number of clusters at time $T$ is $N-T$ for $0 \le T \le N-1$.
The graphical representation of the above procedure on a square lattice is summarized in Fig.~\ref{fig:graphicalrep}.

We show that our rule can describe the random rule and the Achlioptas rule.
Our proposed rule for $q=0$ is equivalent to the random rule.
In the random rule, we randomly choose an edge $e_{ij}$ such that $\tau_{ij}=0$, then we connect the edge $e_{ij}$ ($\tau_{ij}=1$).
In Eq.~(\ref{eq:weight}) for $q=0$, the probabilities are the same $w_{ij}=w_{kl}=1/2$, and thus this situation is equivalent to the random rule.
On the other hand, our rule for $q= +\infty$ realizes the Achlioptas rule.
Hereafter we refer to our rule for $q=-\infty$ as the inverse Achlioptas rule.
In this case we choose to connect the edge where the sum of elements is the largest.
In this way, the network-growth rule can be changed by $q$.
Thus, we can investigate the time evolution of the network depending on various network-growth rules in a unified way.
Notice that because of the conditions imposed in step 2, clusters have no loops, which corresponds to the loopless percolation studied by Manna and Subramanian\cite{Manna-1996}.
%This treatment is the same as the Achlioptas product rule process for square lattice which was considered by Ziff\cite{Ziff-2009}.

\begin{figure}[t]
 \begin{center}
  \includegraphics[scale=0.60]{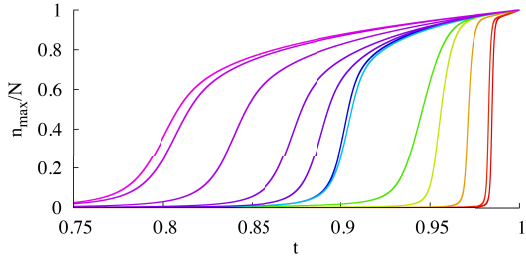}
  \caption{
  (Color online)
  Dynamics of the density of elements in the largest cluster $n_{\rm max}/N$ on a $512 \times 512$ square lattice for $q=-\infty$ (inverse Achlioptas rule), $-1$, $-10^{-1}$, $-10^{-2}$, $-10^{-3}$, $0$ (random rule), $10^{-5}$, $10^{-3}$, $10^{-2}$, $10^{-1}$, $1$, and $+\infty$ (Achlioptas rule) from left to right.
  These results are obtained by averaging out $65536$ independent samples.
  The error bars are smaller than the widths of curves and they are omitted.
  }
  \label{graph:nmax_t}
 \end{center}
\end{figure}

To investigate the network-growth rule dependence of network evolution and the geometric aspects of the percolation cluster at the percolation point, we study the network-growth model on an $L\times L \, (=N)$ two-dimensional square lattice with open boundary conditions.
The coordinate of $v_i$ is represented by the position vector ${\bf r}_i:=(x_i,y_i)$ for $1 \le x_i,y_i \le L$.
In order to understand the $q$-dependence of network evolution quantitatively, we calculate the dynamics of the number of elements in the largest cluster defined by $n_{\rm max}:={\rm max}\{n(\alpha)|1\le \alpha\le N\}$.
The value of $n_{\rm max}$ characterizes the connectivity of the network and is often considered in studies of conventional percolation.
Since clusters are never divided, $n_{\rm max}$ monotonically increases with time.
Figure \ref{graph:nmax_t} shows time development of the density of elements in the largest cluster $n_{\rm max}/N$ for $q=-\infty$ (inverse Achlioptas rule), $-1$, $-10^{-1}$, $-10^{-2}$, $-10^{-3}$, $0$ (random rule), $10^{-5}$, $10^{-3}$, $10^{-2}$, $10^{-1}$, $1$, and $+\infty$ (Achlioptas rule) on a $512\times 512$ square lattice.
These results are obtained by averaging out the results calculated over $65536$ samples.
Here we define $t$ as the normalized time $t:=T/N$.
As shown in Fig.~\ref{graph:nmax_t}, $n_{\rm max}$ becomes explosive as $q$ increases.

\begin{figure}[b]
 \begin{center}
  \includegraphics[scale=0.9]{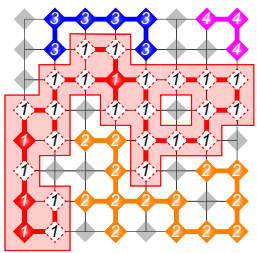}
  \caption{
  (Color online)
  Graphical explanation of percolation.
  The shaded area indicates the percolation cluster.
  The dotted elements denote surface elements.
  In this case, the number of elements in the percolation cluster $n_{\rm p}=26$ and the number of the surface elements in the percolation cluster $n_{\rm s}=22$.
  }
  \label{fig:percolation}
 \end{center}
\end{figure}

\begin{figure}[t]
 \begin{center}
  \includegraphics[scale=0.55]{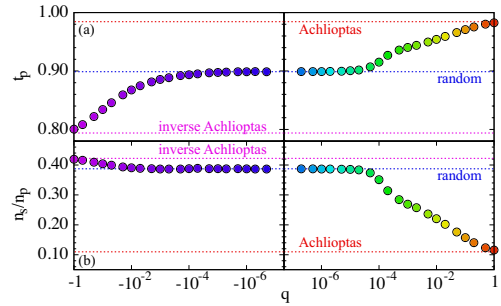}
  \caption{
  (Color online)
  (a) $q$-dependence of the normalized percolation step $t_{\rm p}$ for $L=512$.
  (b) $q$-dependence of the ratio $n_{\rm s}/n_{\rm p}$ for $L=512$.
  These results are obtained by averaging 2048 independent samples.
  The error bars of these results are smaller than the symbol sizes and they are omitted.
  }
  \label{graph:tpL}
 \end{center}
\end{figure}

So far, we constructed a general network-growth rule by introducing a generalized parameter $q$ and showed the $q$-dependence of network evolution.
Next we concentrate on the $q$-dependence of the percolation step.
In this study we define percolation as follows (see Fig.~\ref{fig:percolation}).
If there is a cluster that spreads from the left side to the right side or from top to bottom, we call it a percolation cluster, which is one of its typical definitions in percolation theory.
The percolation step $T_{\rm p}$ is defined by the step when a cluster first becomes percolated.
Hereafter we consider the normalized percolation step: $t_{\rm p}:=T_{\rm p}/N$.
The value of $t_{\rm p}$ is obtained by averaging out the obtained individual percolation steps from $2048$ samples.
Figure \ref{graph:tpL} (a) shows the $q$-dependence of the percolation step for $L=512$.
The percolation step monotonically increases against $q$.
Thus we can change the percolation step by tuning the generalized parameter $q$.
Next we study the $q$-dependence of certain geometric aspects of the percolation clusters.
At first glance, the number of elements $n_{\rm p}$ and surface elements $n_{\rm s}$ in the percolation cluster do not seen to depend on $q$ (not shown).
However we find an obvious relation between $q$ and $n_{\rm s}/n_{\rm p}$ as shown in Fig.~\ref{graph:tpL} (b).
We can study a geometric property of percolation clusters for fixed finite-size systems through $n_{\rm s}/n_{\rm p}$.
When $n_{\rm s}/n_{\rm p}$ is large, the percolation cluster is a porous structure.
In other words, $n_{\rm s}/n_{\rm p}$ indicates the roughness of the percolation cluster.
Figure \ref{graph:tpL} (b) shows the $q$-dependence of the ratio $n_{\rm s}/n_{\rm p}$ for $L=512$.
The ratio $n_{\rm s}/n_{\rm p}$ monotonically decreases against $q$.

\begin{figure*}[t]
 \begin{center}
  \includegraphics[scale=0.84]{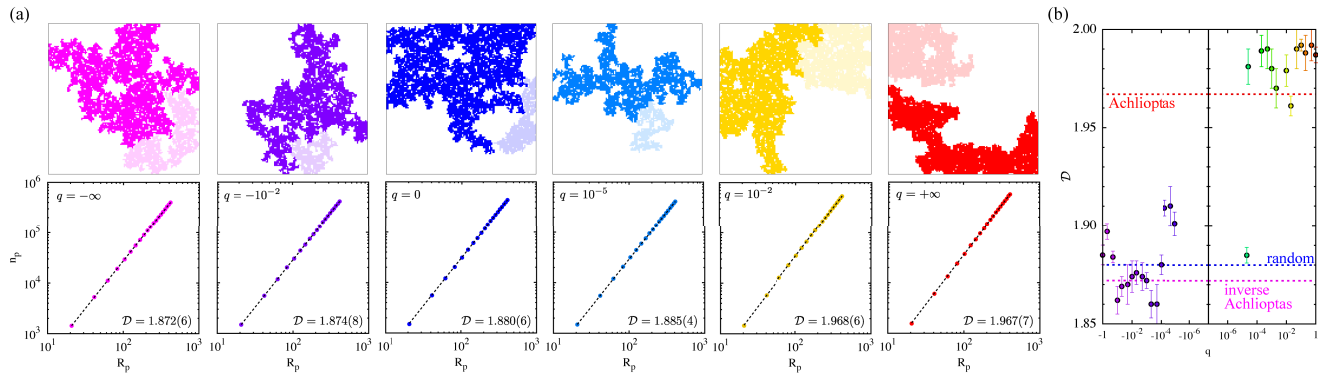}
  \caption{
  (Color online)
  (a) (Upper panels) Snapshots at the percolation point for $q=-\infty$ (inverse Achlioptas rule), $-10^{-2}$, $0$ (random rule), $10^{-5}$, $10^{-2}$, and, $+\infty$ (Achlioptas rule) from left to right.
  The dark and light points depict elements in the percolation cluster and in the second-largest cluster, respectively.
  (Lower panels) Number of elements in the percolation cluster $n_{\rm p}$ as a function of gyradius ${\cal R}_{\rm p}$ for corresponding $q$.
  The dotted lines are obtained by least-squares estimation using Eq.~(\ref{eq:fractal}) and the fractal dimensions ${\cal D}$ are displayed in the bottom right corner.
  (b) $q$-dependence of fractal dimension.
  The dotted lines indicate the fractal dimensions for $q=+\infty$ (Achlioptas rule), $q=0$ (random rule), and $q=-\infty$ (inverse Achlioptas rule) from top to bottom.
  }
  \label{graph:fractal}
 \end{center}
\end{figure*}

In order to investigate the relation between the geometry of a percolation cluster at the percolation point and the network-growth rule in the thermodynamic limit, we consider the $q$-dependence of the fractal dimension.
The fractal dimension can be calculated by the gyradius of the percolation cluster and $n_{\rm p}$.
The gyradius of the $\alpha$-th cluster ${\cal R}(\alpha)$ is defined as
\begin{eqnarray}
 \nonumber
 {\cal R}(\alpha):= \sqrt{\frac{1}{n(\alpha)}\sum_{i\,{\rm s.t.} \sigma_i=\alpha}
  |{\bf r}_i - {\bf r}_0(\alpha)|^2},
 \quad
 {\bf r}_0(\alpha) := \frac{1}{n(\alpha)}\sum_{i\,{\rm s.t.} \sigma_i=\alpha}
  {\bf r}_i,
\end{eqnarray}
where both summations are taken over all the elements such that $\sigma_i=\alpha$, and ${\bf r}_0(\alpha)$ represents the position vector of the center of gravity of the $\alpha$-th cluster.
In this study the definition of the fractal dimension ${\cal D}$ is adopted as follows:
\begin{eqnarray}
 \label{eq:fractal}
 n_{\rm p} \propto {\cal R}_{\rm p}^{\cal D},
\end{eqnarray}
where ${\cal R}_{\rm p}$ denotes the gyradius of the percolation cluster at the percolation point.
The fractal dimension quantitatively characterizes the geometric properties of fractal systems and is often used in the analysis of fractal geometry.
When ${\cal D}=d$, where $d$ is a spatial dimension, the cluster is not fractal.
If $0 < {\cal D} < d$, on the other hand, the geometry of the percolation cluster has a fractal structure as well as a conventional self-similar structure.
The upper panels of Fig.~\ref{graph:fractal} (a) show snapshots of the percolation cluster and the second-largest cluster at the percolation point for $q=-\infty$ (inverse Achlioptas rule), $-10^{-2}$, $0$ (random rule), $10^{-5}$, $10^{-2}$, and, $+\infty$ (Achlioptas rule) from left to right.
The corresponding gyradius dependence of $n_{\rm p}$ is shown in the lower panels of Fig.~\ref{graph:fractal} (a), which are obtained by calculation on lattice sizes from $L=64$ to $L=1280$.
The dotted lines are obtained by least-squares estimation using Eq.~(\ref{eq:fractal}).
Figure \ref{graph:fractal} (b) shows the fractal dimension as a function of $q$.
The fractal dimension for the inverse Achlioptas rule and that for the random rule are of similar value, whereas that for the Achlioptas rule obviously differs from them.
Notice that the fractal dimension of a percolation cluster at the percolation point on a two-dimensional lattice is ${\cal D}=91/48 \simeq 1.896\cdots$\cite{Stauffer-book} which is almost the same value as the fractal dimension for $q=0$.
This result is consistent with the result shown in Fig.~\ref{graph:tpL} (b).
If the cluster is of porous structure, the fractal dimension becomes small.

In this paper, we investigated certain geometric aspects of percolation clusters under a network-growth rule in which the introduced parameter $q$ assigns the rule.
Since our rule includes both the Achlioptas rule\cite{Achlioptas-2009} and the random rule where elements are randomly connected step by step, our rule can be regarded as a generalized network-growth rule.
We studied the time evolution of the number of elements in the largest cluster.
As $q$ increases ({\it i.e.} the rule approaches the Achlioptas rule), the percolation step is delayed and the time evolution of $n_{\rm max}$ becomes explosive.
We also studied another geometric property of the percolation cluster through $n_{\rm s}/n_{\rm p}$, which represents the roughness of the percolation cluster.
The ratio $n_{\rm s}/n_{\rm p}$ monotonically decreases against $q$.
From these facts, the network-growth speed and geometric properties of the percolation cluster change by tuning $q$.
Fractal dimensions for several $q$'s were also calculated.
We found that as $q$ increases, the fractal dimension of percolation cluster increases.
It is expected that the fractal dimension changes continuously as a function of $q$.
However, since the accuracy of the fractal dimension shown in Fig.~\ref{graph:fractal} (b) is not enough to conclude the expectation, we should calculate more larger systems with high accuracy.

In this study we focused on the case of a two-dimensional square lattice.
To investigate the relation between the spatial dimension and more detailed characteristics of percolation ({\it e.g.} critical exponents) for our proposed rule is a remaining problem.
Since our rule is a general rule for many network-growth problems, it enables us to design the nature of percolation.
In this paper, we studied the fixed $q$-dependence of the percolation phenomenon.
However, for instance, in a social network, it is possible that $q$ changes with time.
Then it is an interesting problem to consider the percolation phenomenon with a time-dependent $q$ under our rule.
We strongly believe that our rule will provide a greater understanding of percolation and will be applied for many network-growth phenomena in nature and information technology.

\acknowledgments
The authors would like to thank Takashi Mori and Sergio Andraus for critical reading of the manuscript.
S.T. is partially supported by Grand-in-Aid for JSPS Fellows (23-7601) and R.T. is partially supported by Global COE Program ``the Physical Sciences Frontier'', MEXT, Japan and by National Institute for Materials Science (NIMS).
Numerical calculations were performed on supercomputers at the Institute for Solid State Physics, University of Tokyo.

\end{document}